\documentclass[
reprint,
 aps,
 superscriptaddress,
bibnotes,
pra
]{revtex4-2}
\usepackage{physics}
\usepackage{dsfont}
\usepackage{color,newfloat}
\usepackage{graphicx}
\usepackage{dcolumn}
\usepackage{bm}
\usepackage{amsmath,amssymb,amsthm}
\usepackage{mathtools}
\usepackage{multirow}
\usepackage{hhline}
\usepackage{comment}
\usepackage[dvipsnames]{xcolor}
\usepackage{hyperref}
\usepackage{lineno}
\hypersetup{
    colorlinks=true,
    linkcolor=Maroon,
    citecolor=Maroon,
    urlcolor=Maroon
}
\usepackage{gensymb} 

\usepackage[caption = false]{subfig}
\usepackage{lineno}
\usepackage{siunitx}
\usepackage[mathscr]{euscript}
 \let\mathscr\relax

\usepackage{makecell}

\begin{document}



\title{
Electrically switchable photonic diode empowered by chiral resonance}

\author{Jiaqi Zhao}
\thanks{These authors contributed equally to this work.}
\affiliation{School of Integrated Circuits, Harbin Institute of Technology (Shenzhen), Shenzhen, 518055, China}

\author{Kexun Wu}
\thanks{These authors contributed equally to this work.}
\affiliation{School of Integrated Circuits, Harbin Institute of Technology (Shenzhen), Shenzhen, 518055, China}

\author{Xuecheng Yan}
\affiliation{School of Integrated Circuits, Harbin Institute of Technology (Shenzhen), Shenzhen, 518055, China}

\author{Jin Li}
\affiliation{School of Integrated Circuits, Harbin Institute of Technology (Shenzhen), Shenzhen, 518055, China}

\author{Jiewen Li}
\affiliation{School of Integrated Circuits, Harbin Institute of Technology (Shenzhen), Shenzhen, 518055, China}

\author{Xiaochuan Xu}
\affiliation{School of Integrated Circuits, Harbin Institute of Technology (Shenzhen), Shenzhen, 518055, China}

\author{Ke Xu}
\affiliation{School of Integrated Circuits, Harbin Institute of Technology (Shenzhen), Shenzhen, 518055, China}

\author{Yu Li}
\affiliation{State Key Laboratory of Photonics and Communications, School of Information Science and Electronic Engineering, Shanghai Jiao Tong University, Shanghai, 200240, China}

\author{Linjie Zhou}
\affiliation{State Key Laboratory of Photonics and Communications, School of Information Science and Electronic Engineering, Shanghai Jiao Tong University, Shanghai, 200240, China}

\author{Yan Chen}
\email{chenyan@nudt.edu.cn}
\affiliation{Institute for Quantum Science and Technology, National University of Defense Technology, Changsha, 410073, China}

\author{Jiawei Wang}
\email{wangjw7@hit.edu.cn}
\affiliation{School of Integrated Circuits, Harbin Institute of Technology (Shenzhen), Shenzhen, 518055, China}

\begin{abstract}
On-chip non-reciprocal optical transmission in a magnetic-free route is a key goal for integrated photonics. All-silicon, nonlinearity-enabled photonic diodes present a compelling alternative due to their inherent material compatibility. However, their utility has yet been
constrained by limitations in post-fabrication reconfigurability. 
Here, we present an electrically tunable photonic diode leveraging engineered chiral resonances in an ultra-compact microring architecture. 
The pronounced mode chirality is verified through asymmetric backscattering strengths and interferometric lineshapes in the linear regime. It subsequently enables nonlinearity-driven non-reciprocal transmission, supporting two distinct operational modes with threshold powers as low as $-5\,\mathrm{dBm}$. Moreover, the chiral design grants unprecedented control over self-pulsation dynamics, resulting in oscillation thresholds and temporal signatures determined by the direction of excitation.
Crucially, post-fabrication electrical reconfigurability allows dynamic switching
between forward, backward, and disabled states. This work represents a significant
advancement in integrated non-reciprocal photonics, offering a CMOS-compatible solution
with transformative potential for optical interconnects, photonic neural networks,
and signal processing systems.
\end{abstract}

\date{\today}

\maketitle

Non-reciprocal optical devices are fundamental components of modern optical systems
across classical and quantum domains \cite{bib1,bib2,bib3}. While commercial devices
predominantly rely on magneto-optic (MO) effects  \cite{bib4}, their monolithic integration
remains challenging due to bulky magnets and material incompatibility with the standard
complementary metal-oxide-semiconductor (CMOS) processes \cite{bib5,bib6,bib7}. This limitation has spurred extensive research into
magnetic-free alternatives, including spatiotemporal index modulation  \cite{bib8,bib9} mimicking the MO effect, as well as others circumventing Lorentz reciprocity,
e.g., optomechanical  \cite{bib10,bib11}, optoacoustic  \cite{bib12,bib13}, and nonlinear optical
systems  \cite{bib14,bib15,bib16,bib17,bib18}. Among these, nonlinearity-based approaches
exploit structural and coupling asymmetries to achieve non-reciprocal transmission \cite{bib14,bib19,bib20}. Despite the inherent constraints imposed by dynamic reciprocity  \cite{bib21}, the approach offers great promise for compact, integrated photonic diodes, offering direction-dependent blocking, when simultaneous multiport excitation is not expected \cite{bib22}, making them especially valuable for all-optical shaping and routing of signals in densely integrated systems for optical interconnects  \cite{bib23}, detection  \cite{bib24}, and computing \cite{bib14,bib25}.

Recent advances in chiral optical states within non-Hermitian systems have unlocked
efficient routes for manipulating light flow in photonic integrated circuits (PICs) \cite{bib26}.
In particular, engineered chirality in whispering gallery mode resonators has led to
remarkable phenomena, including asymmetric backscattering  \cite{bib27}, unidirectional
emission  \cite{bib28}, asymmetric mode switching  \cite{bib29}, chiral perfect absorption  \cite{bib30}, chiral light-matter interaction  \cite{bib31},
as well as non-reciprocal transmission  \cite{bib15}.
While the conventional framework of nonlinearity-enabled photonic diodes typically rely on cascaded resonances
\cite{bib14,bib24,bib32} and Fano-type interference \cite{bib33}, chiral resonances emerge as a transformative alternative, delivering on-demand nonreciprocal bandwidth and substantially reduced insertion loss. However, existing implementations remain constrained by their
reliance on static chiral states  \cite{bib23,bib34,bib35}, relying solely on passive nonlinearities
($\chi^{(2)}$ or $\chi^{(3)}$). The inability of electrical reconfigurability in these
photonic diodes represents a key barrier to achieving dynamic control over non-reciprocal
bandwidth, threshold power, and temporal response characteristics.

\begin{figure*}[htbp]
    \centering
    \includegraphics[width=\textwidth]{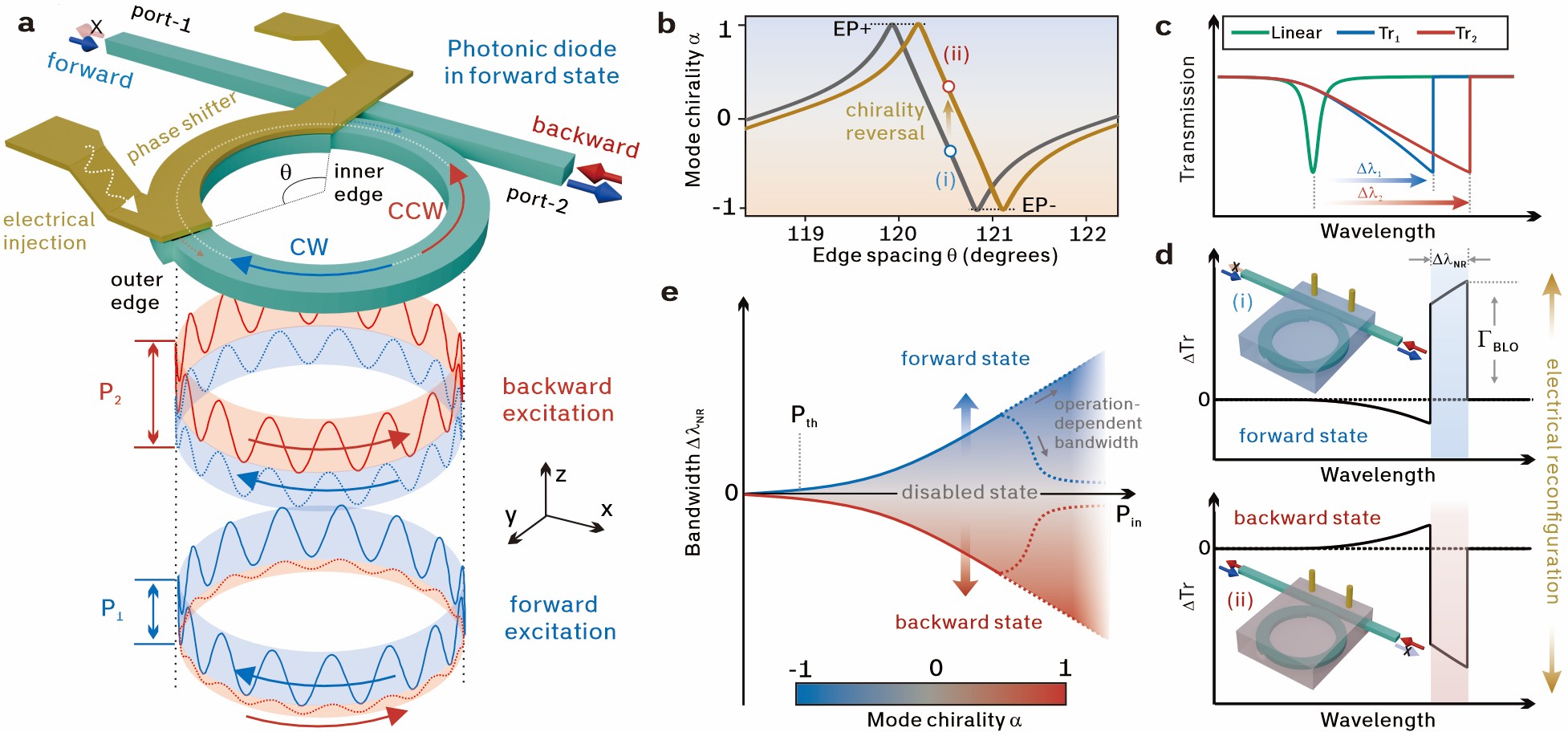}
    \caption{\textbf{Nonreciprocity in a chirality-enabled photonic diode.} 
    \textbf{a}, Schematic of a waveguide-coupled spiral ring on an SOI platform operating as an integrated photonic diode 
    in the forward state ($\alpha < 0$), in which backward transmission is prohibited. 
    The integrated phase shifter enables electrical control of mode chirality. 
    \textbf{b}, Numerically calculated mode chirality $\alpha$ as a function of $\theta$ without (black) 
    and with (brown) electrical tuning. 
    \textbf{c}, Schematic of transmission responses in the linear and nonlinear regimes. 
    The blocking ratio $\Gamma_\text{BLO}$ is defined as the highest achievable transmission contrast within the non-reciprocal band. 
    \textbf{d}, Transmission contrast $\Delta Tr = Tr_{1} - Tr_{2}$ as a function of wavelength, with highlighted reconfigurable 
    non-reciprocal band (shaded areas) in the forward (top) and backward (bottom) states. 
    \textbf{e}, Schematic showing the evolution of the non-reciprocal bandwidth $\Delta \lambda_\text{NR}$ on the injection power $P_\text{in}$ 
    upon tailorable mode chirality, in which three different operational states are denoted.
    The threshold power $P_\text{th}$ is defined as the required injection power to achieve $\Gamma_\text{BLO}$ of $3\,\mathrm{dB}$.}
    \label{fig:fig1}
\end{figure*}

In this article, we propose and experimentally demonstrate an electrically tunable
photonic diode on a silicon-on-insulator (SOI) platform that harnesses chiral
resonances in a compact microring resonator. Driven by unbalanced backscattering
induced by spiral-shaped deformations, strong mode chirality lifts
the contrast in intracavity power between the two excitation directions, turning on the
photonic diode with non-reciprocal transmission at a threshold power as low as $-5\,\mathrm{dBm}$.
The photonic diode operates in two distinct modes, differentiated by thermal pre-activation
conditions, each exhibiting unique spectral characteristics. Besides, the chirality further
governs dynamics of self-pulsation (SP), producing direction-dependent activation thresholds
and waveform characteristics. Crucially, post-fabrication phase tuning permits
full electrical reconfiguration between forward, backward, and disabled states through sign
reversal of chirality. The synergistic combination of electrical reconfiguration and
operation-dependent nonlinear response represents a fundamental advance over conventional
static non-reciprocal devices.


\section*{Results}
\section*{Working principle}

In conventional microring resonators, optical modes exist as degenerate clockwise (CW)
and counterclockwise (CCW) traveling waves. A chiral resonance emerges
through controlled breaking of this degeneracy, typically achieved via either patterned
local scatterers  \cite{bib36,bib37}, or additional mode conversion units  \cite{bib35,bib38,bib39}.
Here our photonic diode employs a compact, holistic design using a microring with
spiral-shaped deformation, combined with integrated phase shifter (Fig.~\ref{fig:fig1}a). 
The azimuthal angle $\theta$ quantifies the angular spacing between the inner and outer spiral edges. The spiral-shaped deformation offers significant flexibility for regulating the non-Hermiticity, through both the inter-modal coupling and external coupling via the continuum (Supplementary Note~1 and Methods) \cite{bib40}. Due to the high-index confinement in this silicon-based structure, the primary effect manifests as two backscattering terms, namely
$\mathbf{A}$ (CW $\rightarrow$ CCW) and $\mathbf{B}$ (CCW $\rightarrow$ CW)  \cite{bib23,bib41}. By varying $\theta$, the mode chirality
$\alpha = (|\mathbf{A}| - |\mathbf{B}|) / (|\mathbf{A}| + |\mathbf{B}|)$
gets efficiently modulated between two extremum cases (see Fig.~\ref{fig:fig1}b), theoretically corresponding to two exceptional points (EPs), namely EP$+$ ($\mathbf{B} = 0, \alpha = 1$) and EP$-$ ($\mathbf{A} = 0, \alpha = -1$),
where the system’s eigenvalues and associated eigenstates coalesce  \cite{bib42,bib43}.

\begin{figure*}[htbp]
    \centering
    \includegraphics[width=\textwidth]{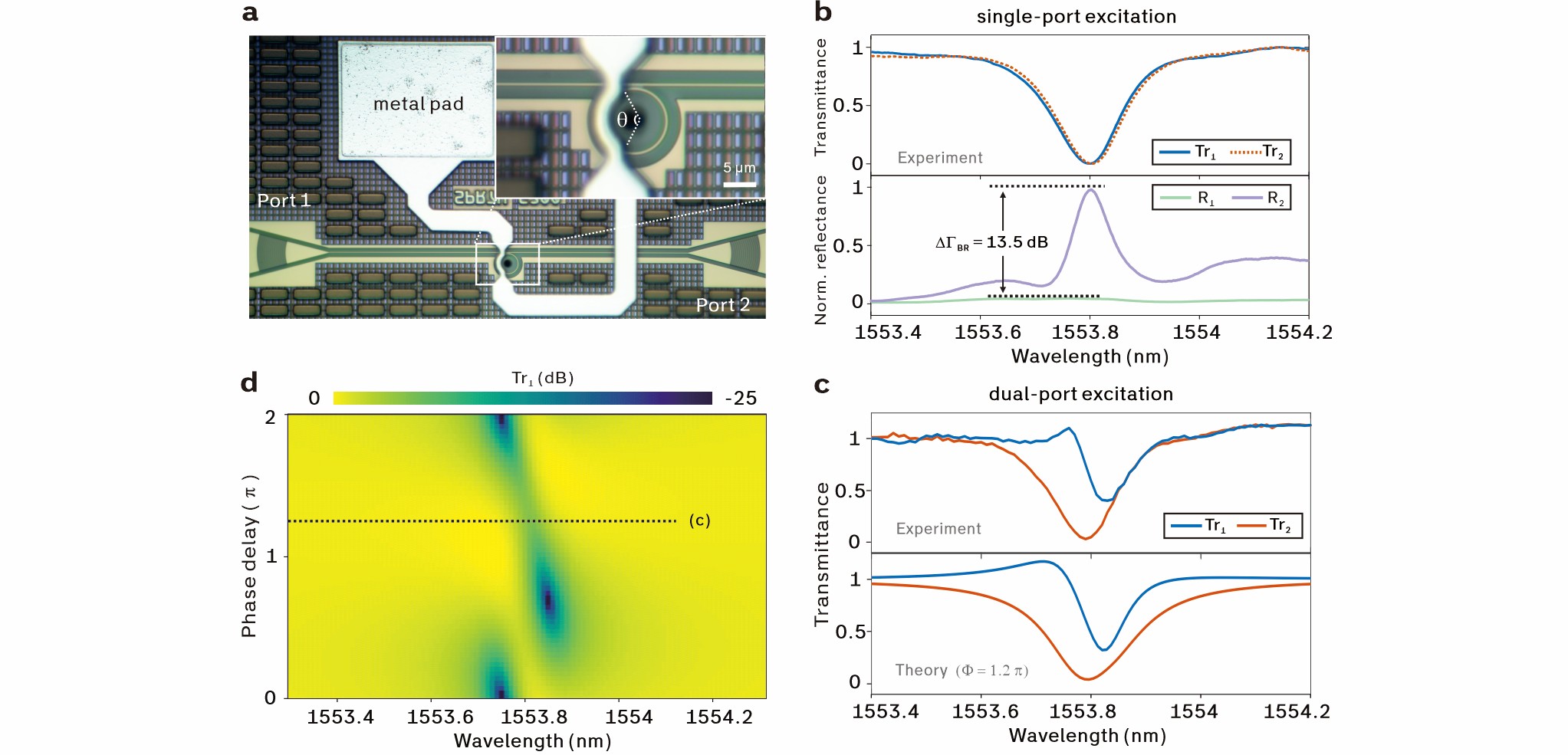}
    \caption{\textbf{Characterization of mode chirality through backscattering and interferometric spectra.} 
    \textbf{a}, Optical microscope image of a waveguide-coupled spiral ring resonator with radius $R$ of $5\,\mu \mathrm{m}$, $\theta = 122.4^\circ$ and deformation parameter $\epsilon = 0.013$. 
    \textbf{b}, Measured transmission spectra $Tr_{1}$ and $Tr_{2}$ (top) and backreflection spectra $R_1$ and $R_2$ (bottom) under single-port excitation. 
    The on-chip injection power $P_\text{in}$ was defined as the optical power coupled into the bus waveguide after the
    grating coupler. Here the estimated $P_\text{in}$ is $-11\,\mathrm{dBm}$.
    \textbf{c},  Transmission spectra $Tr_{1}$ and $Tr_{2}$ (top) under simultaneous dual-port excitation.
    \textbf{d},  Modeled evolution of transmission spectrum $Tr_{1}$ under dual-port excitation with the relative phase delay $ \Phi $ between the two inputs.}
    \label{fig:fig2}
\end{figure*}

In the linear regime, the transmission spectra of this waveguide-coupled ring system under excitation from both ports follow the Lorentz reciprocity theorem (Supplementary Note 2). At high excitation powers, silicon's nonlinear response manifests mainly through the two-photon
absorption (TPA) and the free-carrier absorption (FCA). TPA creates free carriers while
simultaneously introducing heat into the mode volume, while the free-carrier dispersion (FCD)
and thermo-optic (TO) effects induce counteracting resonance shifts (Supplementary Note~3 and Methods).
Overall, the thermal contribution dominates, producing a characteristic redshifted and broadened
resonance  \cite{bib44}. Figure \ref{fig:fig1}c presents the resulting asymmetric lineshape where transmission
abruptly recovers as heating saturates and the resonance returns to its initial wavelength  \cite{bib44}.
The chiral resonance fundamentally alters these nonlinear dynamics for this waveguide-coupled
injection from opposite sides (i.e., ports 1 and 2 in Fig.~\ref{fig:fig1}a). For identical injection power $P_\text{in}$,
asymmetric backscattering creates a strong contrast of the intracavity power, $P_1$ and $P_2$
(see Fig.~\ref{fig:fig1}a), leading to direction-dependent spectral responses. As revealed in the top panel of
Fig.~\ref{fig:fig1}d, a negative chirality ($\alpha < 0$ and $|\mathbf{A}| \ll |\mathbf{B}|$) creates a non-reciprocal
transmission window (the shaded area in Fig.~\ref{fig:fig1}d), establishing the forward state with blocking of backward transmission.

The photonic diode's functionality can be significantly expanded through active control
of chirality. Beyond the fixed geometric parameter $\theta$, the integrated phase shifter enables
adjustment of backscattering strengths via electrical modulation of effective refractive
index change $\Delta n_{\text{eff}}$ (Fig.~\ref{fig:fig1}b)  \cite{bib41}. When chirality is reversed
($\alpha > 0$ and $|\mathbf{A}| \gg |\mathbf{B}|$), the diode switches to its backward state
(the bottom panel of Fig.~\ref{fig:fig1}d). By adjusting $\alpha$ close to near-zero values,
the equalized $P_1$ and $P_2$ effectively deactivate the diode through balanced
absorption-induced broadening (see Fig.~\ref{fig:fig1}e). While both the chirality and the
increased input power generally enhance bandwidth through enhanced nonlinear effects,
we emphasize that the nonlinear scaling of the non-reciprocal bandwidth
$\Delta \lambda_{\text{NR}}$ depends strongly on the specific operation mode, which is
examined quantitatively in the following sections.

\vspace{48pt}

\section*{Non-reciprocity in two operational modes}

\begin{figure*}[htbp]
    \centering
    \includegraphics[width=\textwidth]{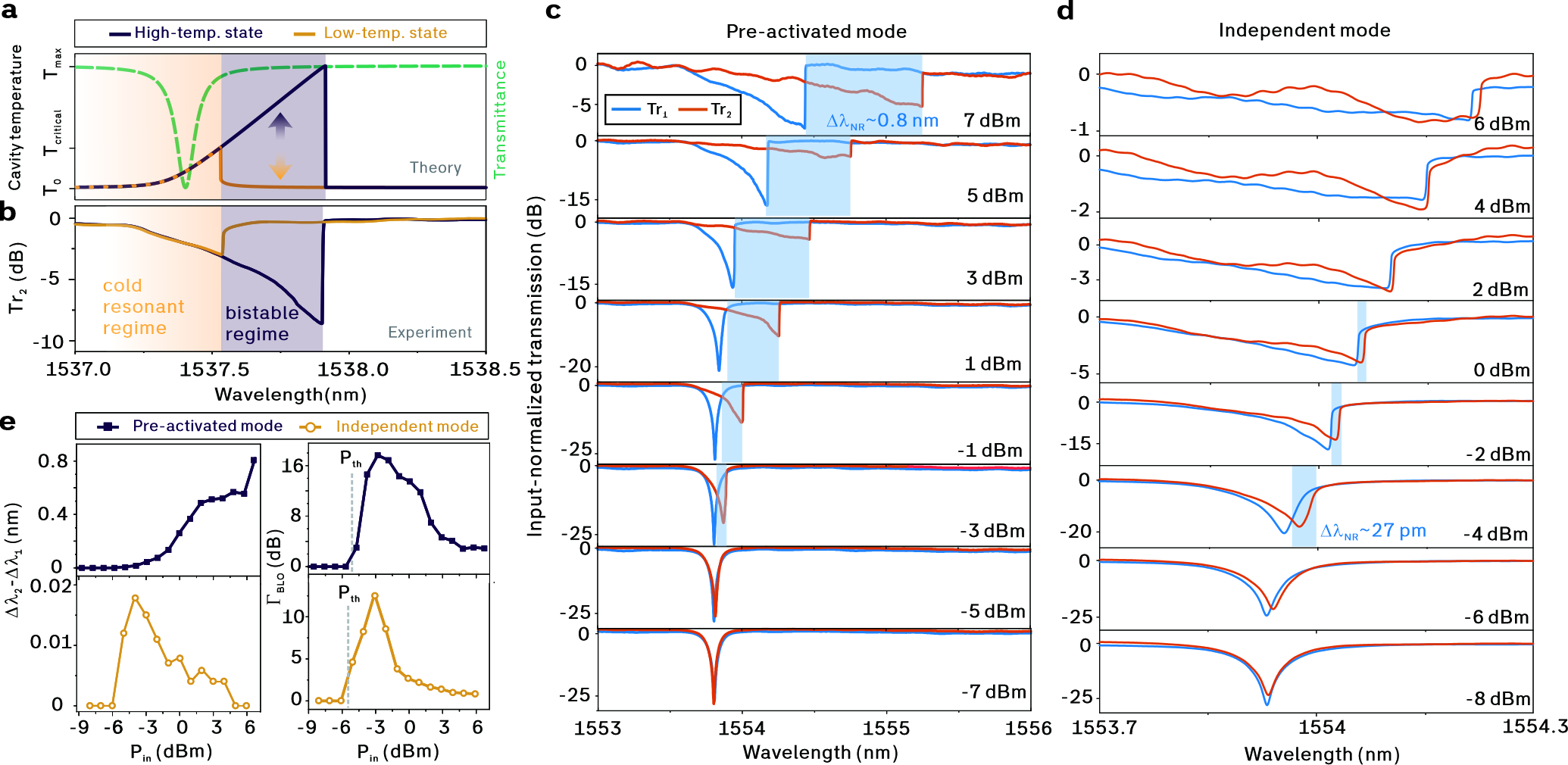}
    \caption{\textbf{Characterization of non-reciprocal transmission in two distinct operation modes.} 
    \textbf{a}, Thermal equilibrium diagram illustrating cavity temperature $T$ versus the probe wavelength for two states. The green curve denotes the transmission spectrum of the cold cavity. 
    The cold resonant regime and bistable regime are denoted with the orange and purple shaded areas, respectively. 
    \textbf{b}, Optical bistability of the system characterized by two transmission spectra $Tr_{2}$ under different laser scanning conditions.
    \textbf{c-d}, Measured transmission spectra $Tr_{1}$ and $Tr_{r}$ for the pre-activated (c) and the independent mode (d), with increasing $P_\text{in}$. 
    The shaded areas represent the non-reciprocal band. 
    \textbf{e}, Summarized evolutions of the relative resonance wavelength offset (left) and $\Gamma_\text{BLO}$ (right) as a function of $P_\text{in}$ for two operation modes. 
    $P_\text{th}$ is denoted with gray dashed lines.}
    \label{fig:fig3}
\end{figure*}

For experimental demonstrations, spiral ring resonators with a compact size were fabricated on an SOI
platform through a foundry service (Fig.~\ref{fig:fig2}a and Methods). As shown in Fig.~\ref{fig:fig2}b, the transmission spectra $Tr_{1}$ (excitation at port 1) and $Tr_{2}$ (excitation at port 2) are characterized upon reversed excitation direction
(Supplementary Note~4 and Methods). For the linear regime, both transmission spectra
which reveal Lorentzian lineshapes, with a loaded Q factor of $\sim 11300$, confirm the
maintenance of Lorentz reciprocity. Meanwhile, the backreflection spectra $R_1$ and $R_2$
revealed pronounced asymmetric backreflection ratio with a contrast $\Delta \Gamma_\text{BR}$
of $\sim 13.5\,\mathrm{dB}$ at the resonance. According to temporal coupled-mode theory (TCMT),
the amplitude of mode chirality of -0.6 can be extracted directly based on the reflection signals, i.e., $\alpha = \frac{\sqrt{R_1} - \sqrt{R_2}}{\sqrt{R_1} + \sqrt{R_2}}$  \cite{bib27}.
To perform a more rigorous estimation of the backscattering coefficients $\mathbf{A}$ and $\mathbf{B}$, we conducted an interferometric experiment by injecting light simultaneously into both ports of the device (Supplementary Note~4). As shown in Fig.~\ref{fig:fig2}c, the spectrum collected at port 1 ($Tr_{2}$) exhibits a standard Lorentzian lineshape. In contrast, the spectrum at port 2 ($Tr_{1}$) displays a distinct Fano-like asymmetry, with the transmission dip shifted spectrally. This asymmetric Fano profile arises from interference between the direct transmission pathway and the backscattered signal, and varies significantly with the relative phase delay $\Phi $ between the two excitation signals (Fig.~\ref{fig:fig2}d). Consequently, this dual-port interferometric technique provides a robust and direct method for extracting the backscattering coefficients $\mathbf{A}$ and $\mathbf{B}$ without relying solely on single-port transmission spectra (Supplementary Note~5).
The numerically modeled spectra for both dual-port (bottom panel, Fig.~\ref{fig:fig2}c) and single-port excitations show excellent agreement with the experimental results.

While entering the nonlinear regime, the photonic diode exhibits two distinct operational modes for non-reciprocal
transmission. The first, termed the ``pre-activated'' mode, employs continuous laser
wavelength up-scanning that progressively heats the cavity, resulting in tracing of the
high-temperature state due to the thermal accumulation and a preserved resonance in
the bistable regime (see Fig.~\ref{fig:fig3}a). This process continues until reaching maximum
temperature $T_{\text{max}}$, where heat accumulation saturates. The second, which we call the
``independent'' mode, employs a pre-defined fixed-wavelength excitation of a cold
resonator. While avoiding cumulative heat accumulation, the system reaches a transient
critical temperature $T_{\text{critical}}$ before entering the bistable regime, and subsequently returning to its
initial temperature $T_0$ along the low-temperature state trajectory (Supplementary Note~6).
Experimental characterization of the transmission spectrum $Tr_{2}$ in Fig.~\ref{fig:fig3}b reveals
that the preactivated mode follows the lower hysteresis trace, while the independent
mode follows the upper hysteresis trace.

The power-dependent characterization reveals striking differences between two operational
modes (Fig.~\ref{fig:fig3}c--d). In the pre-activated mode (Fig.~\ref{fig:fig3}c), asymmetric lineshapes for both
injection directions emerge upon $P_\text{in}$ beyond $-7\,\mathrm{dBm}$. Backward injection generates
larger resonance redshifts compared to forward injection, reaching a maximum $\Gamma_\text{BLO}$ of
$18\,\mathrm{dB}$ at $P_\text{in}$ of $-3\,\mathrm{dBm}$, before degradation due to enhanced nonlinear absorption
effects. The non-reciprocal bandwidth shows continuous expansion with increasing $P_\text{in}$, reaching $\sim 2.9\,\mathrm{nm}$ at $9\,\mathrm{dBm}$,  which represents the upper limit of our characterization system (Supplementary Note~7). Notably, backreflection
measurements confirm these effects (Supplementary Note~8). The backreflection within
the non-reciprocal band upon forward injection becomes much suppressed down to the
noise level, which is close to the ideal scenario of nearly reflectionless transmission
without any resonance-induced perturbation. 

\begin{figure*}[htbp]
    \centering
    \includegraphics[width=\textwidth]{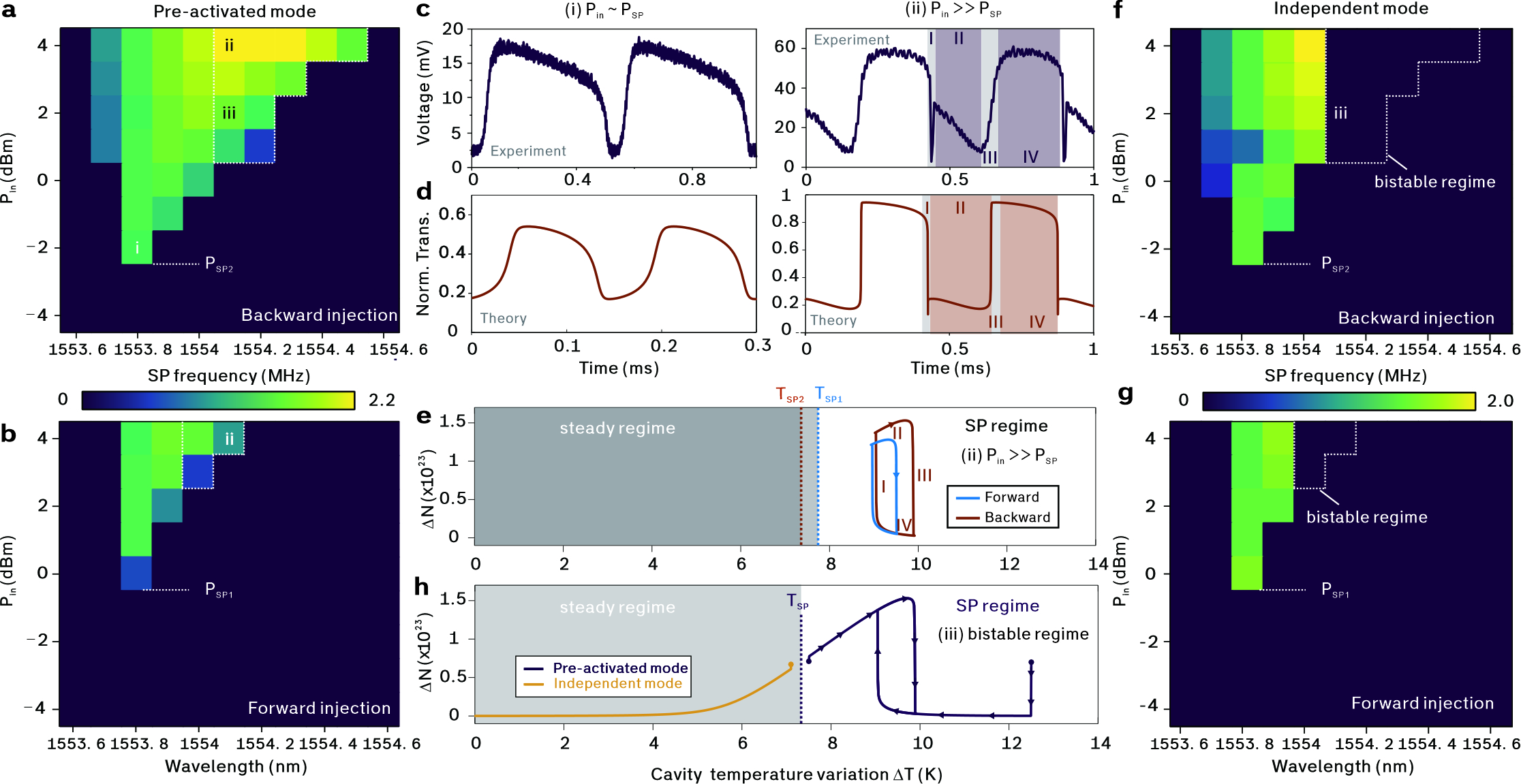}
    \caption{\textbf{Chirality-mediated SP dynamics.} 
    \textbf{a--b}, Maps of SP frequency $\omega_\text{SP}$ as functions of the probe wavelength $\lambda_p$ 
    and the input power $P_\text{in}$ for backward (a) and forward (b) injection performed in the pre-activated mode. 
    \textbf{c}, Measured temporal waveform upon $P_\text{in}$ of $-2\,\mathrm{dBm}$ (left) and $4\,\mathrm{dBm}$ (right) 
    for backward injection. 
    \textbf{d}, Modeled temporal responses for (c). 
    \textbf{e}, Phase diagram showing the trace of $\Delta N$--$\Delta T$ trajectories for two injection directions, 
    with arrows indicating thermodynamic phase progression. 
    \textbf{f--g}, Maps of $\omega_\text{SP}$ as functions of $\lambda_p$ and $P_\text{in}$ for backward (f) 
    and forward (g) injection performed in the independent mode. 
    \textbf{h}, Phase diagram showing the trace of $\Delta N$--$\Delta T$ trajectories upon backward injection 
    in both operation modes.}
    \label{fig:fig4}
\end{figure*}

For the independent mode (Fig.~\ref{fig:fig3}d), the asymmetric lineshapes emerge at a comparable
power level ($P_\text{in} \sim -5\,\mathrm{dBm}$), and the non-reciprocal transmission reaches
$\Gamma_\text{BLO}$ of $\sim 13\,\mathrm{dB}$ upon further  increasing $P_\text{in}$. Notably, the non-reciprocal
bandwidth saturates at $P_\text{in} \sim -4\,\mathrm{dBm}$. This distinction stems from the limit of
resonance redshift that remains constrained within the cold resonant regime (see Fig.~\ref{fig:fig3}e, and Supplementary Note~6). Despite its limited bandwidth, this activation-free “plug-and-play” functionality addresses a key gap in prior implementations and is essential for practical applications with fixed-wavelength lasers.


\section*{Chirality-mediated self-pulsation}

The engineered chiral resonance gives rise to unprecedented control of complex
time-domain dynamics under even a continuous-wave excitation. In experiments at high
$P_\text{in}$ (Fig.~\ref{fig:fig4}a--b), the original equilibrium state becomes unstable through a Hopf bifurcation,
subsequently being re-stabilized as self-sustained pulse sequences with frequency
$\omega_\text{SP}$ around $1$--$2\,\mathrm{MHz}$. The chirality gives rise to markedly distinct SP dynamics, with notably different SP thresholds for opposite injection directions. Forward injection requires higher activation power ($P_\text{SP1} \approx 0\,\mathrm{dBm}$)
to initiate SP compared to backward propagation ($P_\text{SP2} \approx -2\,\mathrm{dBm}$), which is a direct
consequence of the chirality-induced intracavity power imbalance ($P_1 < P_2$). The SP
phenomena evolve dramatically with increasing $P_\text{in}$, with the oscillation band expanding
within the thermally broadened resonance while maintaining distinct directional
characteristics. Forward injection maintains a moderate bandwidth compared to the
backward injection, with nearly reflectionless and lossless transmission outside the SP band.

\begin{figure*}[htbp]
    \centering
    \includegraphics[width=\textwidth]{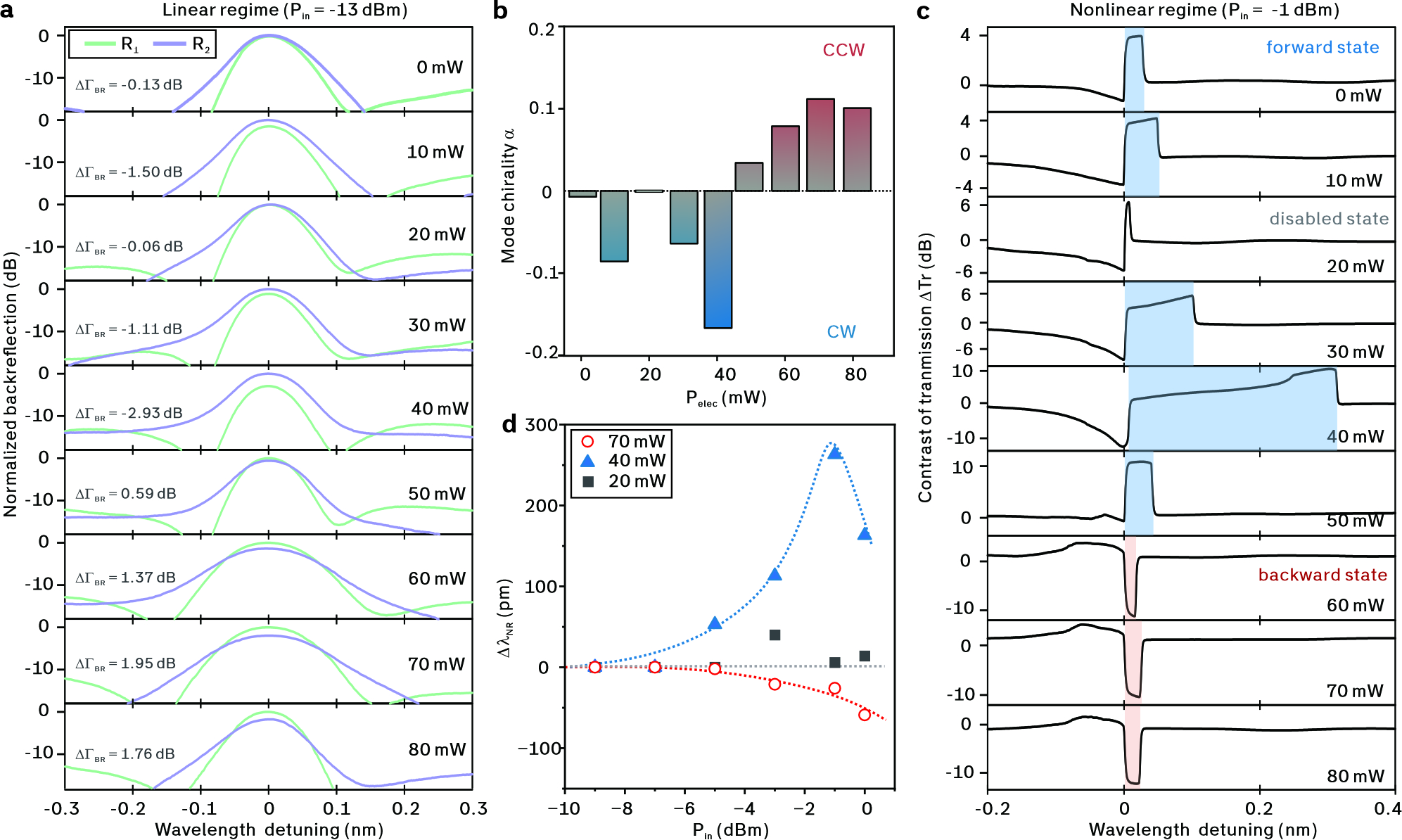}
    \caption{\textbf{Demonstration of an electrically switchable photonic diode.} 
    \textbf{a}, Summarized evolution of backreflection spectra $R_1$ and $R_2$ upon varying $P_\text{elec}$ 
    from $0\,\mathrm{mW}$ to $80\,\mathrm{mW}$, measured at $P_\text{in} = -13\,\mathrm{dBm}$. 
    \textbf{b}, Extracted mode chirality as a function of $P_\text{elec}$. 
    \textbf{c}, Summarized evolution of $\Delta T_r$ upon varying $P_\text{elec}$ from $0\,\mathrm{mW}$ to $80\,\mathrm{mW}$, 
    measured at $P_\text{in} = -1\,\mathrm{dBm}$. 
    \textbf{d}, Summarized evolution of $\Delta \lambda_\text{NR}$ on the injection power $P_\text{in}$ 
    for three representative $P_\text{elec}$ of $20\,\mathrm{mW}$, $40\,\mathrm{mW}$, and $70\,\mathrm{mW}$.}
    \label{fig:fig5}
\end{figure*}

Figure~\ref{fig:fig4}c presents two characteristic cases of SP waveforms. Near the threshold (case~i,
left panel), each oscillation period initiates with a transmission dip ($\sim 0.1\,\mu\mathrm{s}$
duration), followed by a slow decay. For high-power operation (case~ii, right panel), the
oscillating waveform reveals a $\sim 15\,\mathrm{ns}$ transient pulse as a clear signature of the
dominant FCD effect.  These temporal features directly reflect the nonlinear coupling between carrier concentration and temperature, which exhibit distinct dynamic dependencies on the intracavity energy (see Supplementary Note 3). The nonlinear dynamics of our system was numerically modeled (Fig.~\ref{fig:fig4}d), and becomes particularly revealing when analyzed in the phase space defined by free carrier density variation
$\Delta N$ and cavity temperature variation $\Delta T$ (Supplementary Note~9 and Methods).
As $P_\text{in}$ surpasses $P_\text{SP}$, the system undergoes a bifurcation. Beyond this point,
the competing FCD and TO effects can no longer maintain equilibrium, instead forming
stable limit cycles, demonstrating the characteristic Lyapunov stability of self-sustained
oscillations.

Each cycle decomposes into four thermodynamic phases (Fig.~\ref{fig:fig4}e): I) a carrier-driven
phase where FCA induces rapid resonance blueshift through FCD; II) a thermal accumulation
phase dominated by gradual TO-induced redshift; III) a recovery phase as resonance detuning
reduces intracavity power; and IV) a relaxation phase suggesting that the cavity cools back
toward its initial state. Near threshold, weakened FCD effects blend phases I and II,
preventing the system from reaching local equilibrium and producing the observed
transient spikes (Fig.~\ref{fig:fig4}c). The direction-dependent dynamics are confirmed by the distinct trajectories of the limit cycles and the activation temperature $T_{\text{SP2}}$
and $T_{\text{SP1}}$ (case~ii, Fig.~\ref{fig:fig4}e).

The independent mode reveals equally significant but distinct behavior (Fig.~\ref{fig:fig4}f--g).
While maintaining the same directional asymmetry in thresholds, the SP band narrows
substantially, precisely constrained by the bistable regime boundaries. In the bistable
regime, the injection-direction-dependent threshold temperature $T_\text{SPi}$ emerges (Fig.~\ref{fig:fig4}e), which triggers a Hopf bifurcation and guiding the system into limit cycles. Under identical input conditions (e.g., case~iii), the absence of cumulative heating  prevents the resonator from reaching $T_\text{SP}$ for sustained oscillations. Consequently, the system returns directly to the equilibrium state ($\Delta N \approx \Delta T \approx 0$)  instead of following cyclic trajectories (Fig.~\ref{fig:fig4}h). This operational-mode dependence provides an additional degree of freedom over SP characteristics in chirality-engineered photonic systems.

\section*{Electrically switchable photonic diode}

The engineered chiral resonance in our photonic diode presents a generic and robust pathway to on-chip nonreciprocity. Notably, our approach does not require operation at extreme chirality values (i.e., $\pm 1$ at EPs). The full functionality is unleashed through sign-reversible mode chirality, enabling dynamic switching of non-reciprocal transmission direction. To experimentally demonstrate this reconfigurable nonreciprocity, we employ a spiral device with a moderate initial chirality, facilitating its reversal via electrical tuning.
In experiments, the electrical reconfiguration of chirality is first characterized in the
linear regime (Fig.~\ref{fig:fig5}a), where Lorentz reciprocity is well maintained under varying power
applied to the phase shifter $P_\text{elec}$ (Supplementary Note~10). Measurements of the
backreflection spectra $R_1$ and $R_2$ reveal a static $\Delta \Gamma_\text{BR}$ of $\sim 0.13\,\mathrm{dB}$
with zero bias, corresponding to a weak chirality of $\sim -0.007$. As $P_\text{elec}$ increases
from $0$ to $80\,\mathrm{mW}$, active tuning demonstrates remarkable chirality control (Fig.~\ref{fig:fig5}b), with the chirality value tuned to $\sim -0.17$ (at $40\,\mathrm{mW}$) and $\sim 0.11$ (at $70\,\mathrm{mW}$). This demonstrates
in situ reconfiguration of mode chirality without requiring structural modifications.

Nonreciprocity in the nonlinear regime is investigated upon increased $P_\text{elec}$ (Fig.~\ref{fig:fig5}c). Under the static condition, a well-defined non-reciprocal window emerges, with $\Gamma_\text{BLO}$ of $\sim 3.6\,\mathrm{dB}$ and a bandwidth of $\sim 26\,\mathrm{pm}$,
confirming the forward state. At $P_\text{elec} = 40\,\mathrm{mW}$, where chirality reaches
an extremum, the optimal performance yields a maximum $\Gamma_\text{BLO}$ of
$\sim 10.0\,\mathrm{dB}$ and $\Delta \lambda_\text{NR}$ of $\sim 310\,\mathrm{pm}$. As $P_\text{elec}$
further increases to $70\,\mathrm{mW}$, the system exhibits sign-flipped chirality that
inverts the non-reciprocal band with $\Delta \lambda_\text{NR}$ of $\sim 20\,\mathrm{pm}$,
effectively switching the diode to the backward state. Besides, the transition through a
near-neutral chirality ($\sim -0.003$) at $P_\text{elec} = 20\,\mathrm{mW}$ demonstrates
complete nonreciprocity suppression, with residual oscillations showing ultra-narrow
linewidth ($\sim 6\,\mathrm{pm}$) approaching an ideal disabled state. Figure~\ref{fig:fig5}d presents
the evolution of non-reciprocal bandwidth across three characteristic conditions
($P_\text{elec}$ of 20, 40, and 70$\,\mathrm{mW}$), demonstrating the photonic diode's
reconfigurability between forward, backward, and disabled states.

\section*{Discussion}

In summary, we have demonstrated an all-silicon electrically reconfigurable photonic
diode based on spiral-shaped microring resonators with engineered chiral resonances.
Leveraging the sign-reversible mode chirality, the pronounced intracavity power contrast
has enabled magnetic-free non-reciprocal transmission through asymmetric resonance
broadening. Our photonic diodes are benchmarked against existing integrated nonlinearity-enabled nonreciprocal devices on chip (Supplementary Note~11). Featuring a compact cavity footprint of $\sim 100\,\mu\mathrm{m}^2$, our system achieves a record-high nonreciprocal bandwidth of $\sim 2.9\,\mathrm{nm}$ in the pre-activated mode. Notably, we also demonstrate the independent mode, absent in prior chirality-based nonreciprocal systems, exhibiting a bandwidth of $\sim 27\,\mathrm{pm}$. This activation-free “plug-and-play” mode is highly suitable for practical integration with fixed-wavelength lasers.
The strong chirality has not only enabled non-reciprocal transmission but has also provided an
additional pathway to mediate the direction-dependent SP dynamics. Crucially, electrical
reconfiguration of chirality has permitted dynamic switching between forward, backward, and
disabled states.
At this stage, electrical reconfiguration of chirality may require power on the order of tens of milliwatts. This limitation can be effectively addressed through strategies such as integrating micro-scale air trenches for enhanced thermal efficiency or employing non-volatile phase-change materials for electrical trimming. The latter approach, in particular, will enable state retention with zero static power, achieving ultra-low energy-per-reconfiguration and enhanced compatibility with advanced CMOS electronics.

It is important to note that all nonlinearity-enabled photonic diodes are fundamentally bound by dynamic reciprocity and cannot be used as on-chip isolators \cite{bib21}. Nevertheless, our proposed electrically tunable framework opens up new functionalities, with transformative potential, including precision metrology \cite{bib24}, random number generation, and all-optical signal processing. In particular, the reconfigurable chirality offers a new degree of freedom for designing reconfigurable neuromorphic nodes and optical spiking neurons, enabling future neuromorphic \cite{bib45} and edge-computing systems. Beyond these applications, the versatility of reconfigurable chirality is unlocking unforeseen functionalities in quantum photonics, where chiral resonances are emerging as a powerful platform for tailoring quantum emission and cavity quantum electrodynamics \cite{bib46,bib47}.

\section*{Methods}

\subsection*{Modeling}

Our analysis employs an effective non-Hermitian Hamiltonian to describe the chiral
resonances in a deformed microring resonator \cite{bib23,bib48},
\[
H = \begin{pmatrix}
\Omega & \mathbf{A} \\
\mathbf{B} & \Omega
\end{pmatrix}, \tag{1}
\]
where the diagonal elements describe the original mode pair, including the frequency (real
part) and the decay rate (imaginary part), while the off-diagonal elements describe the
backscattering of lightwave from CW to CCW ($\mathbf{A}$) and from CCW to CW ($\mathbf{B}$) direction.  

\begin{figure*}[htbp]
\begin{equation}\label{eq3}
    \frac{\mathrm{d}^2 \Delta T(t)}{\mathrm{d}t^2} + \left( \frac{1}{\tau_{\mathrm{eff}}} \right) \frac{\mathrm{d}\Delta T(t)}{\mathrm{d}t} + \left( \frac{1}{\tau_{\mathrm{s}} \tau_{\mathrm{f}}} \right) \Delta T(t) = \frac{\kappa \Gamma_{\mathrm{Ring}} P_{\mathrm{abs}}(t)}{m c_{\mathrm{p}}} \tag{3}
\end{equation}
\end{figure*}

\begin{figure*}[htbp]
\begin{equation}\label{eq4}
    \frac{\mathrm{d}N(t)}{\mathrm{d}t} = \frac{\Gamma_{\mathrm{FCA}}}{2\hbar \omega_0 n_{\mathrm{g}}^2} \frac{\beta_{\mathrm{Si}} c^2}{V_{\mathrm{FCA}}^2} P_i(t)^2 - \gamma_{\mathrm{fc}} N(t) \tag{4}
\end{equation}
\end{figure*}

The spectral responses can be analyzed via TCMT \cite{bib49} (see Supplementary Note~1--5
for details):
\[
i \frac{d}{dt}
\begin{bmatrix}
\varphi_\text{CCW} \\
\varphi_\text{CW}
\end{bmatrix}
=
H
\begin{bmatrix}
\varphi_\text{CCW} \\
\varphi_\text{CW}
\end{bmatrix}
+ i \sqrt{\kappa_\text{in}}
\begin{bmatrix}
a_{2,\text{in}} \\
a_{1,\text{in}}
\end{bmatrix}. \tag{2}
\]

Where $\kappa_\text{in}$ is the input coupling coefficient, and $a_{1,\text{in}}$ and $a_{2,\text{in}}$
are the excitation amplitudes from ports 1 and 2, respectively. 

The thermal and carrier dynamics within the resonator are governed by equ.~\ref{eq3} and \ref{eq4}. Where $\Delta T$ represents the temperature variation within the mode volume, $\Gamma_{\mathrm{Ring}}$ represents the fractional energy overlap of the mode with the differential temperature within the microring, $m$ is the mass of the resonator, $c_{\mathrm{p}}$ is the specific heat capacity of the resonator. $\tau_{\mathrm{s}}$ and $\tau_{\mathrm{f}}$ stand for the slow and the fast thermal lifetimes ($\tau_{\mathrm{s}} \gg \tau_{\mathrm{f}}$), respectively. The total absorbed optical power converting into heat, $P_{\mathrm{abs}}$, is described by equ.~\ref{eq5},
\[\label{eq5}
P_\text{abs} = (\gamma_\text{lin} + \gamma_\text{TPA} + \gamma_\text{FCA}) \times P_{i}. \tag{5}
\]

Here, $P_i$ denotes the intracavity optical power excited via port $i$ (where $i = 1$ or 2). The terms $\gamma_{\mathrm{lin}}$, $\gamma_{\mathrm{TPA}}$, and $\gamma_{\mathrm{FCA}}$ represent the decay rates associated with linear absorption, TPA and FCA, respectively. The specific formulations for TPA and FCA are given in equ.~\ref{eq6} and \ref{eq7},
\[\label{eq6}
\gamma_{\mathrm{TPA}} = \Gamma_{\mathrm{TPA}} \frac{\beta_{\mathrm{Si}} c^2}{V_{\mathrm{TPA}} n_{\mathrm{g}}^2} P_i, \tag{6}
\]
\[\label{eq7}
\gamma_{\mathrm{FCA}} = c \sigma_{\mathrm{Si}} N / n_{\mathrm{g}}, \tag{7}
\]
where $\Gamma_\mathrm{TPA}$ and $V_\mathrm{TPA}$ denote the confinement factor and the effective mode volume for the respective nonlinear processes. Furthermore, $\beta_{\mathrm{Si}}$ is the TPA coefficient of silicon, $c$ is the speed of light in vacuum, $n_{\mathrm{g}}$ is the group refractive index, $\sigma_{\mathrm{Si}}$ represents the FCA cross-section, and $N$ denotes the free-carrier concentration within the cavity. Further details are provided in Supplementary Note 3.


\subsection*{Device fabrication}

Waveguide-coupled spiral ring resonators were fabricated on an SOI platform via a 
multiple-project wafer (MPW) service (CSiP180A1, CUMEC). The single-mode rib 
waveguide was designed with a thickness $T$ of $220\,\mathrm{nm}$ and a width $W$ of $450\,\mathrm{nm}$. The spiral ring geometry follows the design established in our prior study \cite{bib41}. Here, by optimizing the gap spacing between the waveguide and the microring to be $\sim 200\,\mathrm{nm}$, the microring resonators were designed to work near the critical 
coupling criteria in the linear regime (i.e., the intrinsic loss coefficient is balanced by 
the input-coupling coefficient, $\gamma_{\mathrm{total}} = \kappa_{\mathrm{in}}$). The aluminum-based phase shifter 
was integrated on the segment between the two spiral edges.

\subsection*{Device characterization in the linear regime }

Transmission spectra were characterized using a wavelength-tunable laser source
(TLS570, Santec) at $\sim 1550\,\mathrm{nm}$, coupled into a single-mode fiber through
free-space optics. Transverse-electric (TE) polarization was ensured using a fiber-based
polarization controller (FPC561, Thorlabs). Waveguide-coupled microring resonators were
probed via grating couplers using a dual-fiber configuration, with alignment maintained
by a high-precision positioning system featuring active feedback control (AP-SSAS-SIP-SAXYZ,
Apico). Transmission signals were measured using a high-speed photodetector (HDETIN08-5G-FC,
Lbtek). For characterizing backreflection signals, fibers with facets polished at a tilted angle
of $8^\circ$ (Corning SMF28e, Photostream Technology) together with a circulator
(FCIR-1550-FA, Lbtek) were adopted to suppress background reflections. Spectral acquisition
was performed through wavelength scanning with a step resolution of $1\,\mathrm{pm}$.

\subsection*{Device characterization in the non-linear regime}

Characterizations in the nonlinear regime incorporated excitation using an erbium-doped
fiber amplifier (GA8129-3314, B\&A Technology) with a variable optical attenuator
(FVA-3150, EXFO), offering a tunable range of on-chip injection power $P_\text{in}$
between $-13$ and $9\,\mathrm{dBm}$, with the upper limit constrained by the operational
range of the variable optical attenuator. Non-reciprocal transmission in the pre-activated mode was characterized by
wavelength up-scanning at $100\,\mathrm{nm/s}$. For the independent mode, downward scanning
at the same rate was performed to ensure acquisition efficiency, given that it results in
the same dynamics as stepped scanning. SP waveforms were captured using a high-speed
photodetector and a real-time oscilloscope (MHO5104, Rigol). Electrical reconfiguration
of the photonic diode was implemented through bias voltage application via probe
contacts using a precision source meter (S100, Precise), offering an estimated tuning
efficiency of effective refractive index change of $\sim 2.2 \times 10^{-5}\,\mathrm{RIU}/\mathrm{mW}$.


\section*{Data availability}

The data that support the findings of this study will be made publicly available upon publication of the article.

\section*{Acknowledgement}

We extend our gratitude to Prof. Lorenzo Pavesi and Dr. Stefano Biasi for their fruitful discussions. 
J.W. acknowledges the support from the National Natural Science Foundation of China under Grants 62422503 and 12474375, 
the Guangdong Basic and Applied Basic Research Foundation Regional Joint Fund under Grant 2023A1515011944, 
Science and Technology Innovation Commission of Shenzhen under Grants JCYJ20220531095604009 and RCYX20221008092907027.

\section*{Author contributions}
J. W. and Y. C. conceptualized the project. K. W., J. W. and X. X. performed the theoretical calculations and analysis. K. W., J. W., K. X., Y. L. and L. Z. designed the devices and developed the characterization system. J. Z., X. Y. and J. L. performed the measurements. J. Z., K. W. and Y. C. analyzed the data. J. W., K. W., J. L. and Y. C. wrote the manuscript. All authors discussed the results and contributed to the manuscript.

\noindent\textbf{Notes}
The authors declare no competing financial interest.

\bibliography{main}

\end{document}